\documentstyle[12pt]{article}

\title{Blockspin Cluster Algorithms for Quantum Spin Systems}
\author{U.-J. Wiese and  H.-P. Ying \\[2em]
Institut f\"ur Theoretische Physik, \\
Universit\"at Bern, Sidlerstrasse 5, \\
CH-3012 Bern, Switzerland \\}

\begin{document}
\maketitle
\begin{abstract}

Cluster algorithms are developed for simulating quantum spin systems like
the one- and two-dimensional Heisenberg ferro- and anti-ferromagnets.
The corresponding two- and three-dimensional classical spin models
with four-spin couplings are maped to blockspin models with two-blockspin
interactions. Clusters of blockspins are updated collectively. The efficiency
of the method is investigated in detail for one-dimensional spin chains.
Then in most cases the
new algorithms solve the problems of slowing down from which
standard algorithms are suffering.

\end{abstract}

Two-dimensional quantum spin systems are relevant for the description of the
undoped
anti-ferromagnetic precursor insulators of high-$T_c$ superconductors.
Presumably understanding the physics of the
superconductors requires an understanding of their precursor insulators.
This already is a nontrivial problem, which most likely does not allow for a
complete analytic solution. Therefore it is natural to use a numerical
approach to compute the properties of these materials. At present different
methods are used in numerical studies of quantum spin systems (for a recent
review see for example ref.\cite{Bar91}). Small systems
can be solved completely by a direct diagonalization of the hamiltonian. For
larger systems one can use Monte-Carlo methods. For this purpose the finite
temperature partition function of the $d$-dimensional quantum spin system is
expressed as a pathintegral of a ($d+1)$-dimensional classical system of
Ising-like spin variables with four-spin couplings. The classical system is
then simulated
on a euclidean time lattice with lattice spacing $\epsilon$
using importance sampling techniques like the Metropolis
algorithm. This approach was pioneered by Suzuki and collaborators
\cite{Suz77}. An application to one-dimensional spin chains is described
in ref.\cite{Cul83}. For a simulation of the two-dimensional
Heisenberg anti-ferromagnet see for example ref.\cite{Din90}.
The standard numerical methods are, however,  slowed down in different ways.
First of all most local changes of a configuration are forbidden, because many
configurations have zero Boltzmann weight. Secondly, a change of the
magnetization requires a nonlocal update, which has a very small
acceptance rate at low temperatures. Finally, local algorithms are
critically slowed down in the continuum
limit $\epsilon \rightarrow 0$, because then any correlation length in
euclidean time diverges
in units of $\epsilon$. The slowing down of an algorithm is characterized by
its dynamical critical exponent $z$. The exponent describes how the Monte-Carlo
autocorrelation time $\tau \propto 1/\epsilon^z$ (i.e. the time needed to
create a new statistically independent spin configuration)
behaves in the continuum limit.

For classical spin systems with two-spin couplings --- like the Ising model ---
critical slowing down is almost entirely eliminated by the Swendsen-Wang
\cite{Swe87} and Wolff \cite{Wol89} cluster algorithms for which $z \approx 0$.
Cluster algorithms are, however, not directly applicable to models
with four-spin couplings. In
this paper we map the classical spin models with four-spin couplings to
blockspin models with two-blockspin interactions,
which are then simulated using
the Swendsen-Wang or Wolff cluster algorithms.  The blockspin cluster algorithm
automatically creates allowed spin configurations only. Furthermore, it updates
the magnetization efficiently even at low temperatures.
The efficiency of the blockspin
cluster algorithm is investigated in detail for the one-dimensional spin
chains.
We study the critical slowing down by investigating the
$\epsilon$-dependence of the autocorrelation times of different observables.
For example, the dynamical
exponent of critical slowing down of the algorithm for the one-dimensional
Heisenberg anti-ferromagnet is $z = 0.0(1)$. The use of improved estimators
substantially reduces the variance of measured observables. We should mention
that it is presently not clear how efficient the algorithm works for
two-dimensional spin systems.

We consider quantum systems of spins 1/2 with a Hamilton operator
\begin{equation}
H = J \sum_{x,\mu} \vec{S}_x  \cdot \vec{S}_{x+\hat{\mu}},
\end{equation}
where $\vec{S}_x = \vec{\sigma}_x/2$ is a spin operator located at the
point $x$ of a one-dimensional or two-dimensional quadratic lattice of even
length $L$ with periodic boundary conditions and $\vec{\sigma}_x$ are the
Pauli matrices. The coupling is between nearest neighbors
($\hat{\mu}$ is the unit vector in $\mu$-direction) and $J$ is the exchange
coupling. For $J < 0$ parallel spins are energetically favored and we have a
ferromagnet, while $J > 0$ corresponds to an anti-ferromagnet.

To express the partition function as a pathintegral the hamiltonian of the
one-dimensional spin chain is decomposed into $H = H_1 + H_2$ with
\begin{equation}
H_1 = J
\sum_{x = 2m} \vec{S}_x  \cdot \vec{S}_{x+\hat{1}}, \,\,\,
H_2 = J
\sum_{x = 2m+1} \vec{S}_x  \cdot \vec{S}_{x+\hat{1}}.
\end{equation}
Now one uses the Trotter formula for the partition function
\begin{equation}
Z = \mbox{Tr} \exp(- \beta H) = \mbox{lim}_{N \rightarrow \infty}
\mbox{Tr}[\exp(- \epsilon \beta H_1) \exp(- \epsilon \beta H_2)]^N,
\end{equation}
where $\epsilon = 1/N$ is the lattice spacing in the euclidean time direction
and $\beta$ is the inverse temperature.
The original model is recovered in the continuum limit $\epsilon \rightarrow
0$.
For the two-dimensional system one decomposes $H = H_1 + H_2 + H_3 +
H_4$ with
\begin{eqnarray}
&& H_1 = J \sum_{x = (2m,n)} \vec{S}_x
\cdot \vec{S}_{x+\hat{1}}, \,\,\,
H_2 = J
\sum_{x = (2m+1,n)} \vec{S}_x  \cdot \vec{S}_{x+\hat{1}},
\nonumber \\
&& H_3 = J \sum_{x = (m,2n)} \vec{S}_x
\cdot \vec{S}_{x+\hat{2}}, \,\,\,
H_4 = J
\sum_{x = (m,2n+1)} \vec{S}_x  \cdot \vec{S}_{x+\hat{2}},
\end{eqnarray}
and one uses the Suzuki formula
\begin{equation}
Z = \mbox{lim}_{N \rightarrow \infty}
\mbox{Tr}[\exp(- \epsilon \beta H_1) \exp(- \epsilon \beta H_2)
\exp(- \epsilon \beta H_3) \exp(- \epsilon \beta H_4)]^N.
\end{equation}
In the next step the partition function is expressed as a pathintegral of
Ising-like variables by inserting complete sets of eigenstates
$|1\rangle$ and $|\mbox{--1}\rangle$ of $\sigma_x^3$
between the factors $\exp(- \epsilon \beta H_i)$. This gives rise to a
$(d+1)$-dimensional classical spin system with periodic boundary conditions in
the euclidean time direction. The $H_i$ are sums of
commuting terms. The nonzero elements of the transfer matrix are given by
\begin{eqnarray}
&& \langle \mbox{1 1}|\exp(- \epsilon \beta
J \vec{S}_x \cdot \vec{S}_{x+\hat{\mu}})
|\mbox{1 1}\rangle =
\langle \mbox{--1 --1}|\exp(- \epsilon \beta
J \vec{S}_x \cdot \vec{S}_{x+\hat{\mu}})
|\mbox{--1 --1}\rangle = \nonumber \\
&& \exp(- \epsilon \beta J/4), \nonumber \\
&& \langle \mbox{1 --1}|\exp(- \epsilon \beta
J \vec{S}_x \cdot \vec{S}_{x+\hat{\mu}})
|\mbox{1 --1}\rangle =
\langle \mbox{--1 1}|\exp(- \epsilon \beta
J \vec{S}_x \cdot \vec{S}_{x+\hat{\mu}})
|\mbox{--1 1} \rangle = \nonumber \\
&& \exp(- \epsilon \beta J/4) \frac{1}{2}(1 + \exp(\epsilon \beta J)),
\nonumber \\
&& \langle \mbox{1 --1}|\exp(- \epsilon \beta
J \vec{S}_x \cdot \vec{S}_{x+\hat{\mu }})
|\mbox{--1 1}\rangle =
\langle \mbox{--1 1}|\exp(- \epsilon \beta
J \vec{S}_x \cdot \vec{S}_{x+\hat{\mu}})
|\mbox{1 --1}\rangle =  \nonumber \\
&& \exp(- \epsilon \beta J/4) \frac{1}{2}(1 - \exp(\epsilon \beta J)).
\end{eqnarray}
All other elements are equal to zero.
For a ferromagnet ($J < 0$) the expressions are positive and can hence be
interpreted as Boltzmann factors
\begin{equation}
\langle s_1 s_2|\exp(- \epsilon \beta
J \vec{S}_x \cdot \vec{S}_{x+\hat{\mu}})|s_3 s_4 \rangle =
\exp(- S[s_1,s_2,s_3,s_4])
\end{equation}
of spin configurations $s = \pm 1$ with a classical
euclidean action $S$. For an anti-ferromagnet, on the other hand,
this is not possible
because some transfer matrix elements are negative. Therefore in the
anti-ferromagnetic case one performs a unitary transformation of the original
hamiltonian by rotating every second spin by an angle $\pi$.
The nonzero elements of the transfer matrix are then positive for $J > 0$.
Expressed as a pathintegral the partition function takes the form
\begin{equation}
Z = \prod_{x,t} \sum_{s(x,t) = \pm 1} \exp(- S)
\end{equation}
with
\begin{eqnarray}
\exp(- S) & = &
\prod_{x=2m,t=2p} \exp(- S[s(x,t),s(x+\hat{1},t),s(x,t+1),s(x+\hat{1},t+1)])
\nonumber \\ & \times &
\prod_{x=2m+1,t=2p+1} \exp(-
S[s(x,t),s(x+\hat{1},t),s(x,t+1),s(x+\hat{1},t+1)])
\nonumber \\
\end{eqnarray}
in the one-dimensional case and with
\begin{eqnarray}
\exp(- S) & = &
\prod_{x=(2m,n),t=4p} \exp(-
S[s(x,t),s(x+\hat{1},t),s(x,t+1),s(x+\hat{1},t+1)])
\nonumber \\ & \times &
\prod_{x=(2m+1,n),t=4p+1}
\exp(- S[s(x,t),s(x+\hat{1},t),s(x,t+1),s(x+\hat{1},t+1)])
\nonumber \\ & \times &
\prod_{x=(m,2n),t=4p+2}
\exp(- S[s(x,t),s(x+\hat{2},t),s(x,t+1),s(x+\hat{2},t+1)])
\nonumber \\ & \times &
\prod_{x=(m,2n+1),t=4p+3}
\exp(- S[s(x,t),s(x+\hat{2},t),s(x,t+1),s(x+\hat{2},t+1)]) \nonumber \\
\label{Boltz2d}
\end{eqnarray}
for the two-dimensional system.
Note that the classical spins $s(x,t)$ interact with each
other via four-spin couplings. Most spin configurations are
forbidden (their Boltzmann factor vanishes) because the corresponding elements
of the transfer matrix are zero. This is a problem
for standard local algorithms because most local changes of a configuration
are not allowed.
Therefore it is natural to attempt a collective nonlocal update
of the spins. For spin models with two-spin couplings this can be done using
the Swendsen-Wang \cite{Swe87} or Wolff \cite{Wol89} cluster algorithms, which
flip whole clusters of spins simultaneously. The cluster algorithms can,
however, not be applied directly to models with four-spin interactions.

To make an application of cluster algorithms possible we map the classical spin
models with four-spin couplings to blockspin models with two-blockspin
couplings. A blockspin is a collection of a few spins. For
the one-dimensional spin chain we define blockspins consisting of four spins
located at the corners of a plaquette
\begin{equation}
b(2m,2p) = \{s(x,t),s(x+\hat{1},t),s(x,t+1),s(x+\hat{1},t+1)\}
\end{equation}
with $x = 2m-1$ and $t=2p$.
Note that  each spin belongs to exactly one blockspin and
the blockspins live one a lattice with a doubled lattice spacing.
The original action determines the action for the blockspins. For example, the
four-spin interactions for $x=2m,t=2p$ induce couplings between space-like
nearest neighbor blockspins at $(2m,2p)$ and $(2m+2,2p)$
\begin{equation}
S[b(2m,2p),b(2m+2,2p)] = S[s(x,t),s(x+\hat{1},t),s(x,t+1),s(x+\hat{1},t+1)],
\end{equation}
and the four-spin interactions for $x=2m+1,t=2p+1$ induce couplings between
time-like nearest neighbor blockspins at $(2m,2p)$ and $(2m,2p+2)$
\begin{equation}
S[b(2m,2p),b(2m,2p+2)] = S[s(x,t),s(x+\hat{1},t),s(x,t+1),s(x+\hat{1},t+1)].
\end{equation}
The spins can also be arranged to blockspins in another way
\begin{equation}
\tilde{b}(2m,2p) = \{s(x,t),s(x+\hat{1},t),s(x,t+1),s(x+\hat{1},t+1)\}
\end{equation}
where now $x=2m$ and $t=2p-1$.
Then the four-spin interactions for $x=2m,t=2p$ induce two-blockspin couplings
between time-like nearest neighbors and the interactions at $x=2m+1,t=2p+1$
induce couplings between space-like
nearest neighbors. The two blocking schemes of spins into
blockspins are illustrated in figs.1a,b. To ensure ergodicity
an updating algorithm for the blockspins
must alternate between the two blocking schemes.

\begin{figure}
\caption{Three different blocking schemes. The blockspins $b$ and $\tilde{b}$
consist of four spins and
live on lattices with a doubled lattice spacing. The crossed plaquettes carry
the four-spin interaction which turns into a two-blockspin interaction. In the
third scheme one blockspin $b_0$ consists of all spins with the same space
coordinate $x_0$. \newline $\;$}
\begin{picture}(120,120)
\put(0,0){\line(1,0){120}}
\put(0,20){\line(1,0){120}}
\put(0,40){\line(1,0){120}}
\put(0,60){\line(1,0){120}}
\put(0,80){\line(1,0){120}}
\put(0,100){\line(1,0){120}}
\put(0,120){\line(1,0){120}}
\put(0,0){\line(0,1){120}}
\put(20,0){\line(0,1){120}}
\put(40,0){\line(0,1){120}}
\put(60,0){\line(0,1){120}}
\put(80,0){\line(0,1){120}}
\put(100,0){\line(0,1){120}}
\put(120,0){\line(0,1){120}}
\put(0,80){\line(1,1){40}}
\put(0,40){\line(1,1){80}}
\put(0,0){\line(1,1){120}}
\put(40,0){\line(1,1){80}}
\put(80,0){\line(1,1){40}}
\put(20,0){\line(-1,1){20}}
\put(60,0){\line(-1,1){60}}
\put(100,0){\line(-1,1){100}}
\put(120,20){\line(-1,1){100}}
\put(120,60){\line(-1,1){60}}
\put(120,100){\line(-1,1){20}}
\put(28,6) {$b$} \put(68,6) {$b$} \put(108,6) {$b$}
\put(28,46) {$b$} \put(68,46) {$b$} \put(108,46) {$b$}
\put(28,86) {$b$} \put(68,86) {$b$} \put(108,86) {$b$}
\end{picture}
$\; \; \; \; \; \; \;$
\begin{picture}(120,120)
\put(0,0){\line(1,0){120}}
\put(0,20){\line(1,0){120}}
\put(0,40){\line(1,0){120}}
\put(0,60){\line(1,0){120}}
\put(0,80){\line(1,0){120}}
\put(0,100){\line(1,0){120}}
\put(0,120){\line(1,0){120}}
\put(0,0){\line(0,1){120}}
\put(20,0){\line(0,1){120}}
\put(40,0){\line(0,1){120}}
\put(60,0){\line(0,1){120}}
\put(80,0){\line(0,1){120}}
\put(100,0){\line(0,1){120}}
\put(120,0){\line(0,1){120}}
\put(0,80){\line(1,1){40}}
\put(0,40){\line(1,1){80}}
\put(0,0){\line(1,1){120}}
\put(40,0){\line(1,1){80}}
\put(80,0){\line(1,1){40}}
\put(20,0){\line(-1,1){20}}
\put(60,0){\line(-1,1){60}}
\put(100,0){\line(-1,1){100}}
\put(120,20){\line(-1,1){100}}
\put(120,60){\line(-1,1){60}}
\put(120,100){\line(-1,1){20}}
\put(8,26) {$\tilde{b}$} \put(48,26) {$\tilde{b}$} \put(88,26) {$\tilde{b}$}
\put(8,66) {$\tilde{b}$} \put(48,66) {$\tilde{b}$} \put(88,66) {$\tilde{b}$}
\put(8,106) {$\tilde{b}$} \put(48,106) {$\tilde{b}$} \put(88,106) {$\tilde{b}$}
\end{picture}
$\; \; \; \; \; \; \;$
\begin{picture}(120,120)
\put(0,0){\line(1,0){120}}
\put(0,20){\line(1,0){120}}
\put(0,40){\line(1,0){120}}
\put(0,60){\line(1,0){120}}
\put(0,80){\line(1,0){120}}
\put(0,100){\line(1,0){120}}
\put(0,120){\line(1,0){120}}
\put(0,0){\line(0,1){120}}
\put(20,0){\line(0,1){120}}
\put(40,0){\line(0,1){120}}
\put(59.5,0){\line(0,1){120}}
\put(59.6,0){\line(0,1){120}}
\put(59.7,0){\line(0,1){120}}
\put(59.8,0){\line(0,1){120}}
\put(59.9,0){\line(0,1){120}}
\put(60,0){\line(0,1){120}}
\put(60.1,0){\line(0,1){120}}
\put(60.2,0){\line(0,1){120}}
\put(60.3,0){\line(0,1){120}}
\put(60.4,0){\line(0,1){120}}
\put(60.5,0){\line(0,1){120}}
\put(80,0){\line(0,1){120}}
\put(100,0){\line(0,1){120}}
\put(120,0){\line(0,1){120}}
\put(0,80){\line(1,1){40}}
\put(0,40){\line(1,1){80}}
\put(0,0){\line(1,1){120}}
\put(40,0){\line(1,1){80}}
\put(80,0){\line(1,1){40}}
\put(20,0){\line(-1,1){20}}
\put(60,0){\line(-1,1){60}}
\put(100,0){\line(-1,1){100}}
\put(120,20){\line(-1,1){100}}
\put(120,60){\line(-1,1){60}}
\put(120,100){\line(-1,1){20}}
\put(28,6) {$b$} \put(88,26) {$\tilde{b}$}
\put(28,46) {$b$} \put(88,66) {$\tilde{b}$}
\put(28,86) {$b$} \put(88,106) {$\tilde{b}$} \put(62,6) {$b_0$}
\end{picture}
\end{figure}
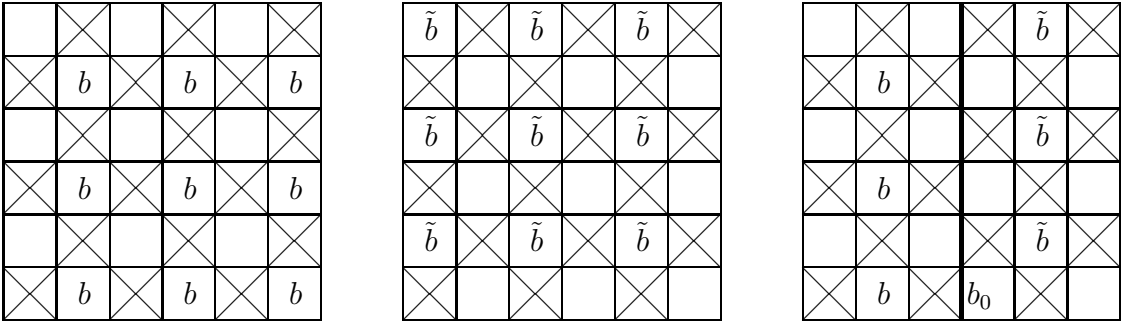

For the two-dimensional system a blockspin
consists of eight spins located at the corners of a cube
\begin{eqnarray}
&& b(2m,2n,2p) =
\{s(x,t),s(x+\hat{1},t),s(x+\hat{2},t),s(x+\hat{1}+\hat{2},t), \nonumber \\
&& s(x,t+1),s(x+\hat{1},t+1),s(x+\hat{2},t+1),s(x+\hat{1}+\hat{2},t+1)\}
\end{eqnarray}
with $x=(2m-1,2n-1)$ and $t=2p$. Again the blockspin lattice has a doubled
lattice spacing. Another blocking scheme is given by
\begin{eqnarray}
&& \tilde{b}(2m,2n,2p) =
\{s(x,t),s(x+\hat{1},t),s(x+\hat{2},t),s(x+\hat{1}+\hat{2},t), \nonumber \\
&& s(x,t+1),s(x+\hat{1},t+1),s(x+\hat{2},t+1),s(x+\hat{1}+\hat{2},t+1)\}
\end{eqnarray}
where now $x = (2m,2n)$ and $t = 2p-1$.
Using the action of eq.(\ref{Boltz2d}) it is straightforward to
show that both schemes induce two-blockspin interactions only.
Again, to ensure ergodicity the algorithm should alternate between the two
schemes.

The cluster algorithms make use of a flip symmetry of the blockspin model.
A blockspin $b = \{s_1,s_2,...,s_n\}$ is flipped to $-b =
\{-s_1,-s_2,...,-s_n\}$ simply by flipping all
spins that belong to $b$. It is clear that the blockspin action is
invariant against flipping all blockspins
simply because the original action is invariant against flipping all
spins. The cluster algorithm puts bonds between nearest neighbor blockspins
$b$ and $b'$ with the probability
\begin{equation}
p = 1 - \mbox{min}\{1,\exp(- S[-b,b'])/\exp(- S[b,b'])\}.
\end{equation}
Two blockspins which are connected by a bond belong to the same cluster. The
algorithm flips all blockspins in one cluster simultaneously.
In the Swendsen-Wang multi-cluster method each cluster is flipped with the
probability 1/2. In the Wolff single-cluster method one blockspin is randomly
selected and the cluster to which it belongs is flipped with probability 1.
Recently, it has been realized that the single-cluster algorithm can be
vectorized \cite{Eve91}.
Using $S[b,-b'] = S[-b,b']$ and $S[-b,-b'] = S[b,b']$ one can show that both
algorithms obey detailed balance. The blockspin cluster algorithm
automatically generates allowed spin configurations only. If one attempts to
flip $b$ when $\exp(- S[-b,b']) = 0$, i.e. when the new configuration
is forbidden, the algorithm puts a bond with probability
$p = 1 - \mbox{min}\{1,0\} = 1$. Hence $b$ and $b'$ fall in the same cluster
and
must be flipped simultaneously. The cluster growth comes to an end only
when the new configuration is allowed. When one uses the single-cluster
method it is guaranteed that one generates a new
different configuration in each step.

As defined up to now the blockspin cluster algorithms are not ergodic because
they change the magnetization $M = 1/2 \sum_x s(x,t_0)$ and the staggered
magnetization $M_s = 1/2 \sum_x (-1)^x s(x,t_0)$ by even numbers only.
$M$ and $M_s$ are integers (rather than half-integers)
because we work on a lattice with an even number of
space points. Note that $M$ is independent of $t_0$ because the transfer matrix
commutes with the total spin. To allow $M$ and $M_s$ to change by odd numbers
also, we
introduce extra blocking schemes with some block spins $b_0 =
\{s(x_0,t)  \; \mbox{for all} \; t \}$
consisting of all spins with the same space coordinates $x_0$.
The other blockspins are taken from the above schemes.
A typical example is shown in fig.1c. Similarily, we include blocking
schemes which allow the so-called
winding number $N_w = 1/2 \sum_t s(x_0,t)$ to change by
an odd integer.

The Swendsen-Wang and Wolff cluster algorithms eliminate critical slowing down
from the Ising model because it is not frustrated. For strongly
frustrated models,
for example for spin glasses, cluster algorithms do not work efficiently.
Therefore the question arises if our blockspin models are frustrated or not.
To answer this question for the spin chains one must investigate all allowed
configurations of four nearest neighbor blockspins located at the corners of
a square of the blockspin lattice. Nearest neighbor pairs of blockspins
interact via bonds (the four-spin interactions of the original
classical spin model). Now the Boltzmann factor of a bond is compared
to the value it has when one of the two blockspins at its ends
is flipped. The blockspin
configuration with the larger Boltzmann factor is the favored configuration of
this bond. If one can flip the four blockspins such that all four
bonds between them are in their favored configuration the configuration
is not frustrated. We have verified that the blockspin model
for the one-dimensional anti-ferromagnet has indeed no frustrated
configurations. For the
one-dimensional ferromagnet, on the other hand, some configurations are
frustrated. In a frustrated configuration one of the four bonds
cannot be in its favored configuration. However, one can show that the
Boltzmann factor of the disfavored bond is in all cases only a factor
$(1 + \exp(\epsilon \beta J))/2$ smaller than  the favored Boltzmann factor.
In the continuum limit $\epsilon \rightarrow 0$ the factor goes to 1 and the
frustration disappears. A priori it is not clear if a weak frustration
can slow down the cluster algorithm in the ferromagnetic case. As we will see
later, the blockspin cluster algorithm does in fact work perfectly only in the
anti-ferromagnetic case. However, also for the ferromagnet the autocorrelation
times of the cluster algorithm are much smaller
than the ones of the Metropolis algorithm. For the two-dimensional
spin systems both for the ferromagnet and for the anti-ferromagnet some
blockspin configurations are frustrated. In addition, not all
frustrations disappear in the continuum limit. Therefore, one
should not expect that the blockspin cluster algorithm for two-dimensional
quantum spin systems works as well as in the one-dimensional case. However,
we expect that it still decorrelates faster than standard algorithms.

Cluster algorithms offer the possibility to use improved estimators which
reduce
the variance of different observables. For example, for the single-cluster
method the susceptibility can be expressed as
\begin{equation}
\chi = \frac{\beta}{L^d} \langle M^2 \rangle =
2dN \beta \langle \frac{M_{\cal C}^2}{|{\cal C}|}\rangle,
\end{equation}
where $L^d$ is the spatial volume,
$2dN$ is the number of points in the euclidean time direction,
$|{\cal C}| = \sum_{(x,t) \in {\cal C}} 1$ is the cluster size and
$M_{\cal C} = 1/2 \sum_{(x,t_0) \in {\cal C}} s(x,t_0)$
is the cluster magnetization. It is interesting to note that also $M_{\cal C}$
is independent of $t_0$. As a consequence, clusters with nonzero magnetization
must wrap around the lattice in the euclidean time direction. Small clusters
which do not wrap around the lattice
have $M_{\cal C} = 0$. Let $L_{\cal C}^d$ be the
minimal spatial extent of a cluster closed in euclidean time, i.e.
$L_{\cal C}^d = \mbox{min}_{t_0}\{\sum_{(x,t_0) \in {\cal C}} 1\}$.
The cluster magnetization is then
limited by $|M_{\cal C}| \leq L_{\cal C}^d/2$ and the cluster
size is restricted by $|{\cal C}| \geq 2dN L_{\cal C}^d$ such that
\begin{equation}
\langle L_{\cal C}^d \rangle \geq \frac{4 \chi}{\beta}.
\end{equation}
The algorithm updates the susceptibility efficiently
only if many clusters wrapping around the euclidean time direction
fit into the lattice. This requires $\langle L_{\cal C}^d \rangle \ll L^d$.
Hence we expect that the cluster algorithm works efficiently only if
\begin{equation}
\chi \ll \frac{\beta L^d}{4}.
\end{equation}
The ferromagnetic systems have degenerate ground states with total
spin $L^d/2$. In the zero temperature limit the susceptibility is then given by
$\chi = \beta (L^d+2)/12$ and the algorithm does not work efficiently.
In all other
cases, for example at non-zero temperature or for anti-ferromagnets, $\chi$
stays finite as the spatial volume goes to infinity and the algorithm
should work. One can also
define an improved estimator for the staggered susceptibility
$\chi_s = \beta \langle M_s^2 \rangle/L^d$. Since the staggered magnetization
is
$t_0$-dependent also small clusters contribute to $\chi_s$.
Finally, we introduce the internal
energy density $e = - 1/L^d (d \; \mbox{ln} Z / d\beta)$.

Now let us turn to the numerical results. We have tested the blockspin cluster
algorithm in detail for one-dimensional spin chains. To test the
efficiency of the blockspin cluster algorithm we compare it to a Metropolis
update of the blockspins. In fact, the blockspin Metropolis algorithm is
similar to the standard algorithms commonly used.
For both algorithms we
measure the autocorrelations of the susceptibility
\begin{equation}
C_{\chi}(\delta\tau) = \overline{\chi(\tau) \chi(\tau + \delta\tau)}
\end{equation}
as a function of the Monte-Carlo time
difference $\delta\tau$. From this we obtain the integrated
autocorrelation time $\tau_{\chi}$ as
\begin{equation}
\exp(- 1/\tau_{\chi}) = \sum_{\delta\tau = 1}^{\infty} C_{\chi}(\delta\tau) /
\sum_{\delta\tau = 0}^{\infty} C_{\chi}(\delta\tau).
\end{equation}
Note that $\tau_{\chi} = \tau_0$ for $C_{\chi}(\delta\tau) \propto
\exp(- \delta\tau/\tau_0)$.
In the same way we define the integrated autocorrelation times $\tau_{\chi_s}$
of the staggered susceptibility and
$\tau_e$ of the internal energy density. For the cluster algorithm we use the
single-cluster method and we measure the autocorrelations based on the improved
estimators. Our results are summarized in table 1.
The quoted autocorrelation
times are in units of Monte-Carlo sweeps. A sweep is defined such that it takes
about the same CPU-time for the cluster and for the Metropolis algorithm.
In all cases we have performed a random start followed by
5000 sweeps for thermalization and by 50000 sweeps for measurements.

\begin{table} \begin{center}
\caption{Results for the one-dimensional Heisenberg spin chains.
The blockspin single-cluster algorithm (C) is compared
to the Metropolis algorithm (M). $\newline \;$}
\begin{tabular}{|c|c|c|c|c|c|c|c|c|c|c|c|} \hline
A & $J$ & $\beta$ & $L$ & $2N$ & $\chi$ & $\tau_{\chi}$ & $\chi_s$ &
$\tau_{\chi_s}$ & $e$ & $\tau_e$ \\ \hline
M & -1 & 1 & 32 &  32 & 0.352(8)  & 8.1(5)  & 0.176(2)  & 1.8(2)
& -0.1347(6) & 1.7(2)  \\ \hline
C & -1 & 1 & 32 &  32 & 0.364(2)  & 0.52(3) & 0.1796(6) & 0.52(3)
& -0.1328(5) & 1.4(1)  \\ \hline
M & -1 & 1 & 32 &  64 & 0.378(8)  & 9.0(5)  & 0.180(2)  & 2.4(2)
& -0.1335(7) & 3.0(2)  \\ \hline
C & -1 & 1 & 32 &  64 & 0.371(2)  & 0.58(3) & 0.1790(6) & 0.48(3)
& -0.1338(5) & 1.5(1)  \\ \hline
M & -1 & 1 & 32 & 128 & 0.348(8)  & 8.0(3)  & 0.179(2)  & 2.6(2)
& -0.133(1)  & 5.6(3)  \\ \hline
C & -1 & 1 & 32 & 128 & 0.369(1)  & 0.55(3) & 0.1790(6) & 0.49(2)
& -0.1338(5) & 1.5(1)  \\ \hline
M & -1 & 1 & 32 & 256 & 0.377(8)  & 8.5(3)  & 0.177(2)  & 3.3(2)
& -0.133(1)  & 10.7(4) \\ \hline
C & -1 & 1 & 32 & 256 & 0.369(1)  & 0.56(3) & 0.1782(6) & 0.54(3)
& -0.1339(5) & 1.5(1)  \\ \hline
M & -1 & 2  & 128 &  16 & 0.80(7)   & 94(8)     & 0.332(1)  & 1.0(1)
& -0.1878(2) & 2.4(2) \\ \hline
C & -1 & 2  & 128 &  16 & 0.935(2)  & 0.56(2)   & 0.3304(9) & 0.49(2)
& -0.1881(2) & 1.8(1) \\ \hline
M & -1 & 4  & 128 &  32 & 2.4(4)    & 180(15)   & 0.665(6)  & 4.9(3)
& -0.2212(2) & 4.9(3) \\ \hline
C & -1 & 4  & 128 &  32 & 2.52(1)   & 1.2(1)    & 0.664(1)  & 0.9(1)
& -0.2211(1) & 1.9(1) \\ \hline
M & -1 & 8  & 128 &  64 & 8(4)      & 3300(200) & 1.29(1)   & 22(3)
& -0.2379(2)  & 5.4(6) \\ \hline
C & -1 & 8  & 128 &  64 & 7.30(8)   & 3.9(3)    & 1.331(5)  &  2.9(2)
& -0.23744(9) & 1.6(1) \\ \hline
C & -1 & 16 & 128 &  16 & 19.5(3)   & 5.1(2)    & 2.58(1)   & 3.9(2)
& -0.24680(2) & 1.4(1) \\ \hline
C & -1 & 16 & 128 &  32 & 20.6(4)   & 9.5(8)    & 2.68(2)   & 9.9(9)
& -0.24550(3) & 1.2(1) \\ \hline
C & -1 & 16 & 128 &  64 & 22.6(6)   & 18(2)     & 2.67(2)   & 17(2)
& -0.24482(5) & 1.7(1) \\ \hline
C & -1 & 16 & 128 & 128 & 23.2(7)   & 17(2)     & 2.64(3)   & 19(2)
& -0.24475(8) & 1.9(2) \\ \hline
M &  1 & 1 & 32 &  32 & 0.134(2)  & 4.0(2)  & 0.426(6)  & 4.3(2)
& -0.207(1)  & 2.8(1)  \\ \hline
C &  1 & 1 & 32 &  32 & 0.1354(4) & 0.34(3) & 0.4275(9) & 0.43(4)
& -0.2049(8) & 1.8(1)  \\ \hline
M &  1 & 1 & 32 &  64 & 0.137(2)  & 3.7(1)  & 0.429(6)  & 5.1(2)
& -0.205(1)  & 4.9(2) \\ \hline
C &  1 & 1 & 32 &  64 & 0.1362(4) & 0.42(3) & 0.426(1)  & 0.45(3)
& -0.2029(8) & 2.0(1) \\ \hline
M &  1 & 1 & 32 & 128 & 0.140(2)  & 3.9(2)  & 0.422(7)  & 5.9(2)
& -0.203(2)  & 7.3(3) \\ \hline
C &  1 & 1 & 32 & 128 & 0.1374(4) & 0.41(3) & 0.4246(9) & 0.41(3)
& -0.2036(8) & 1.9(1) \\ \hline
M &  1 & 1 & 32 & 256 & 0.138(2)  & 3.4(3)  & 0.404(7)  & 5.5(3)
& -0.204(2)  & 13(1)  \\ \hline
C &  1 & 1 & 32 & 256 & 0.1362(4) & 0.39(2) & 0.4246(9) & 0.44(3)
& -0.2038(8) & 1.8(1) \\ \hline
M &  1 & 2  & 128 &  16 & 0.148(5)  & 17(1)     & 1.24(1)   & 1.5(1)
& -0.3443(4) & 1.9(1) \\ \hline
C &  1 & 2  & 128 &  16 & 0.1435(6) & 0.28(2)   & 1.239(2)  & 0.41(2)
& -0.3443(4) & 2.2(1) \\ \hline
M &  1 & 4  & 128 &  32 & 0.122(5)  & 23(1)     & 3.34(3)   & 2.5(1)
& -0.4226(2)  & 2.0(1) \\ \hline
C &  1 & 4  & 128 &  32 & 0.1251(5) & 0.37(2)   & 3.399(7)  & 0.41(2)
& -0.4217(2)  & 1.6(1) \\ \hline
M &  1 & 8  & 128 &  64 & 0.14(2)   & 200(20)   & 8.5(1)    & 12(1)
& -0.4410(2)  & 2.2(2) \\ \hline
C &  1 & 8  & 128 &  64 & 0.1173(5) & 0.42(2)   & 8.51(2)   & 0.49(3)
& -0.4410(1)  & 1.3(1) \\ \hline
C &  1 & 16 & 128 &  16 & 0.0714(6) & 0.92(5)   & 47.5(1)   & 0.70(4)
& -0.59088(6) & 0.85(5) \\ \hline
C &  1 & 16 & 128 &  32 & 0.0985(6) & 0.59(2)   & 25.9(1)   & 0.61(2)
& -0.48977(4) & 0.99(5) \\ \hline
C &  1 & 16 & 128 &  64 & 0.1087(7) & 0.58(5)   & 21.59(5)  & 0.56(5)
& -0.45460(6) & 1.00(5) \\ \hline
C &  1 & 16 & 128 & 128 & 0.1120(7) & 0.65(5)   & 20.76(5)  & 0.61(4)
& -0.44512(7) & 1.04(9) \\ \hline
\end{tabular} \end{center} \end{table}

The Metropolis algorithm can change the magnetization only by flipping a
blockspin $b_0$ which consists of all spins with the same space coordinate
$x_0$. The flip is allowed only if all spins of $b_0$ are parallel.
Surprisingly, the probability for this situation
--- and hence the acceptance rate of
the global step --- is finite in the continuum limit. Therefore the
autocorrelation times $\tau_{\chi}$ and $\tau_{\chi_s}$ do not diverge in the
continuum limit. However, they are at least an order of magnitude larger than
the ones of the cluster algorithm.
The autocorrelation time of the internal energy density, on
the other hand, diverges as $\tau_e \propto 1/\epsilon^{z_e}$.
For the Metropolis
algorithm our $\beta = 1$ data both for the ferro- and for the
anti-ferromagnet yield a dynamical critical exponent $z_e = 0.8(1)$
while the cluster algorithm has $z_e = 0.0(1)$. Note that the exponent may
depend on the observable considered,
because it was extracted from the integrated
autocorrelation time and not from the exponential decay of an autocorrelation
function.

Although the global step of the Metropolis algorithm
is not critically slowed down
(its acceptance rate is $\epsilon$-independent) it is very severely
slowed down at low
temperatures (large $\beta$). This prevents the application
of standard numerical methods to quantum spin systems at low temperatures.
Already at $\beta = 8$ the Metropolis algorithm has $\tau_{\chi} = 3300(200)$
for the ferromagnet and $\tau_{\chi} = 200(20)$ for the anti-ferromagnet.
The cluster algorithm, on the
other hand, has autocorrelation times of at most a few sweeps and one can go
to lower temperatures like $\beta = 16$ very easily. For the ferromagnet the
inequality $\chi \ll \beta L^d/4$ puts a lower limit on the temperature when
the spatial volume is fixed. Our low temperature data for the ferromagnet
at $\beta = 16$ show some
slowing down but the autocorrelation times are moderate.
For the anti-ferromagnet, on the other hand, there is no
indication of slowing down when the temperature is lowered or when the
continuum limit is approached and $z_{\chi} = z_{\chi_s} = z_e = 0.0(1)$.

To summarize we have developed blockspin cluster algorithms for one- and
two-dimensional quantum spin systems. Our results for spin chains
show that the new algorithms eliminate slowing down for anti-ferromagnets.
For ferromagnets slowing down is also practically eliminated as long as the
temperature is not too small. One should, however, be careful in generalizing
our findings to two-dimensional quantum spin
systems. Since the corresponding blockspin models
have some frustrated configurations, it is presently
not clear how efficient the
algorithm works in this case. A detailed analysis of the efficiency
of the blockspin cluster algorithm for the two-dimensional Heisenberg
anti-ferromagnet is in progress. We like to mention that blockspin cluster
algorithms can also be applied to other models. First of all the algorithm also
works for spin systems with anisotropic couplings or (with slight
modifications) in an external magnetic field. Secondly, it can be applied
to models of higher spins ($S = 1,3/2,...$) if it is combined with a method
that changes the absolute value of the spin projection $|S_x^3|$.
The one-dimensional
Heisenberg spin chain is equivalent to a six-vertex model which is a special
eight-vertex model. General eight-vertex models with a
spin flip symmetry can also be treated with blockspin cluster algorithms.
Finally, the algorithm may be useful for simulations of one-dimensional
Fermi-systems like the ones described in ref.\cite{Hir82}.

It is a pleasure
to thank P. Hasenfratz who initiated our interest in quantum spin
systems for his support.


\begin{thebibliography}{8}
\bibitem{Bar91}
T. Barnes, Int. J. Mod. Phys. C2 (1991) 659.
\bibitem{Suz77}
M. Suzuki, Prog. Theor. Phys. 56 (1976) 1454; \\
M. Suzuki, S. Miyashita and A. Kuroda, Prog. Theor. Phys. 58 (1977) 1377.
\bibitem{Cul83}
J. J. Cullen and D. P. Landau, Phys. Rev. B27 (1983) 297.
\bibitem{Din90}
M. S. Makivi\'{c} and H.-Q. Ding, Phys. Rev. B43 (1991) 3562.
\bibitem{Swe87}
R. Swendsen and J.-S. Wang, Phys. Rev. Lett. 58 (1987) 86.
\bibitem{Wol89}
U. Wolff, Phys. Rev. Lett. 62 (1989) 361.
\bibitem{Eve91}
H. G. Evertz, FSU-SCRI-91-183 (1991).
\bibitem{Hir82}
J. E. Hirsch, R. L. Sugar, D. J. Scalapino and R. Blankenbecler,
Phys. Rev. B26 (1982) 5033.
\end{thebibliography}
\end{document}